\documentclass[aps, prapplied, preprint]{revtex4-2}

\usepackage{booktabs}
\usepackage{braket}
\usepackage{amsmath}
\usepackage{graphicx}
\usepackage[font=small]{caption}
\usepackage{subcaption}
\usepackage{array}
\usepackage{placeins}
\usepackage{multirow}
\usepackage{tabularx}
\usepackage{tablefootnote}
\usepackage{adjustbox}
\usepackage{hyperref}
\graphicspath{{./}}

\makeatletter
\def\@email#1#2{%
 \endgroup
 \patchcmd{\titleblock@produce}
  {\frontmatter@RRAPformat}
  {\frontmatter@RRAPformat{\produce@RRAP{*#1\href{mailto:#2}{#2}}}\frontmatter@RRAPformat}
  {}{}
}%
\makeatother

\begin{document}

\title[Phase diagram and electronic structure of {KCuTe_{1-x}Se_{x}} \ldots]{First principles phase diagram calculation and theoretical investigation of electronic structure properties of $\mathrm{KCuTe_{1-x}Se_{x}}$ for photocathode applications}

\author{Arini Kar}
\author{K.R. Balasubramaniam}
\author{Dayadeep S. Monder}
 
\affiliation{ 
Department of Energy Science and Engineering, Indian Institute of Technology Bombay, Mumbai, Maharashtra 400076}
 \email{dmonder@ese.iitb.ac.in}

\date{\today}

\begin{abstract}
Recent high-throughput studies of copper-based semiconductors have identified potassium-based copper chalcogenides, KCuA (A $\in$ Te, Se, S) as optimal light absorbers in photovoltaic and photoelectrochemical devices. In this work, we investigate the applicability of $\mathrm{KCuTe_{1-x}Se_{x}}$ as photocathode materials using first-principles calculations. The calculated temperature-composition phase diagram predicts the formation of the solid solution of $\mathrm{KCuTe_{1-x}Se_{x}}$ in the hexagonal structure beyond a maximum critical temperature of 322 K. Structure relaxation in the alloy competes with volume deformation of the parent lattice and charge exchange between (Te, Se) anions to produce a net linear variation in the bandgap with the alloy concentration. The following results suggest that this alloy is a suitable photocathode: i) lower effective mass, and hence a higher mobility of electrons in the alloy compared to the end compounds, ii) an absorption coefficient of the order of $\mathrm{10^{5}\ cm^{-1}}$, and iii) an appropriate alignment of the conduction band with respect to the hydrogen reduction reaction.
\end{abstract}

\maketitle

\baselineskip=\normalbaselineskip

\section{\label{sec:intro} Introduction}
Photocathodes that can be paired with particular photoanodes in high-efficiency tandem photoelectrochemical devices remain an active area of research due to the stringent material requirements. A low band gap, the appropriate alignment of the conduction band with the proton reduction reaction, and stability in an aqueous medium are essential conditions for the application of semiconductors as photocathodes \citep{Sivula2011}. The performance of a tandem photoelectrochemical (PEC) device with a given photoanode can be further optimized by tuning the bandgap of the photocathode while retaining its other desirable properties to ensure complementarity of the bandgaps of the two photoelectrodes \citep{Lewis2013}.

Highly mismatched alloys are widely used to tune the bandgap of photocathodes and solar absorbers \citep{KANESHIRO201012, Zhou2018, JANG2025129, Ye2020, Butson2019, Weyers1992, Shan1999, Wu2002, Alberi2007, Yu2016, Jaquez2015, Broesler2009, Mayer2010, Ting2015, Ting2019, Wang2011, MANSOUR2018, Bar2004, Chen2011, Fan2013}. Broadly, two types of effects on the bandgap upon alloying are observed, depending on the material system. Linear variation (Vegard law behavior) of the bandgap with alloy concentration is observed in kesterite and chalcopyrite solid  solutions \citep{KANESHIRO201012, Zhou2018, JANG2025129, Wang2011, MANSOUR2018, Bar2004, Chen2011, Fan2013}. In the other extreme, a complex band bowing behavior, attributable to the band anti-crossing interaction (BAC), is observed in some II-VI and III-V semiconductor alloys \citep{Ye2020, Butson2019, Weyers1992, Shan1999, Wu2002, Alberi2007, Yu2016, Jaquez2015, Broesler2009, Mayer2010, Ting2015, Ting2019}. Importantly, such alloy systems exhibit significant reduction in the band gap, i.e. $>$ 1.0 eV reduction compared to the parent, at very low alloy concentrations, making these extremely stable wide band gap materials viable options for visible light photocatalysis. For example, the bandgap of ZnO reduces from 3.3 eV to 2.1 eV when alloyed with 1.2\% Se at the anion site \citep{Mayer2010}. A subset of these HMAs exhibit remarkable photocurrents and solar-to-hydrogen (STH) efficiencies and are promising candidates for photocathodes \citep{KANESHIRO201012, Zhou2018, JANG2025129, Ye2020, Butson2019}. However, intrinsic defects and toxic, scarce, and expensive constituents of these materials hinder/limit their performance and scalability.
  
A recent high-throughput study on Cu-based semiconductors identified KCuTe as a potential absorber material for solar cells due to defect tolerance, which leads to high carrier lifetime along with the presence of non-toxic and earth-abundant elements in the compound \citep{Dahliah2021}. Another first-principles study has proposed the entire class of potassium-based ternary copper chalcogenides, KCuA (A $\in$ S, Se, Te), as light absorbers and photocathodes. The predicted bandgaps of KCuSe and KCuTe are 1.28 eV and 1.62 eV \citep{Behera2024} respectively. Indeed, KCuTe and KCuSe have been experimentally synthesized in the hexagonal structure \citep{Savelsberg1978}. We suggest that the band gap of the KCuTe photocathode can be modulated by alloying with large-size and chemically mismatched Se atoms at the anion site to optimize PEC efficiency.

Since only solid solutions of HMAs present the desired variation in the band gap with alloy concentration, thermodynamic analysis of the stable phases of alloys often precedes investigations into their electronic structure properties \citep{Biswas2008, BREIDI2009, ANNANE2010, Fan_2009, ZHU20171, Long2017, Long2018386, Kar2023, Ouendadji2011, YunbinHe2018CdSeS, Chen2011, Matsubara2022, LI201596, TAOXUE2014251}. A recent thermodynamic study of $\mathrm{ZnO_{1-x}Se_{x}}$ alloy particularly emphasizes the importance of such thermodynamic phase diagrams \citep{Kar2023}. The study shows that $\mathrm{ZnO_{1-x}Se_{x}}$ cannot form a solid solution in the entire composition range; the alloy can form terminal solutions at the O-rich end up to $x \leq$ 0.05 in the wurtzite structure, and in the sphalerite structure at the Se-rich end when 0.5 $\leq x \leq$ 1. Only these terminal solid solutions exhibit extreme bandgap reduction.

In this work, we have computed and studied the temperature-composition phase diagram of $\mathrm{KCuTe_{1-x}Se_{x}}$ using both the solid solution model and CE-based MC simulations. We have elucidated the contributions of volume deformation, charge exchange, and structure relaxation in the lattice to the energetics of the alloy. The solid solution of the stable phases determined from the phase diagram is used to study the electronic band structure and the optical properties of the alloys. We have further investigated the applicability of the alloys as photocathodes on the basis of the effective mass of charge carriers and band alignment with the water redox potential.

\section{\label{sec:method} Computational details}
The temperature-composition phase diagram of the alloys are calculated from the solid solution model. The solid solutions of $\mathrm{KCuTe_{1-x}Se_{x}}$ are modeled using special quasirandom structures (SQS) of 48 atoms that mimic the random alloy up to 20 pairs of nearest neighbors \citep{Wei1990}. The energy of formation of the SQS are given by Equation \ref{eq:eof}.

\begin{eqnarray}\label{eq:eof}
\Delta E(x) =&& E(KCuTe_{1-x}Se_{x}) \nonumber \\
&&- [(1-x)E(KCuTe) + xE(KCuSe)]
\end{eqnarray}

Here, $x$ is the alloy concentration and $E(KCuTe_{1-x}Se_{x})$, $E(KCuTe)$ and $E(KCuSe)$ are the energies of SQS, pure KCuTe and pure KCuSe, respectively, calculated via DFT as implemented in Quantum Espresso \citep{Giannozzi_2017}. The electronic core is approximated with the norm-conserving pseudopotential and the exchange-correlation functional is approximated using Perdew-Burke-Ernzerhof (PBE) \citep{Perdew1996}. The energy cut-off is set to 80 Ry. The convergence threshold for the self-consistent field (SCF) calculation is $10^{-10}$ Ry. The energy and force convergence threshold for the ionic relaxation step are $10^{-6}$ Ry and $10^{-4}$ a.u., respectively. 1600 K points per reciprocal atom (KPPRA) are used for $\mathrm{KCuTe_{1-x}Se_{x}}$ in the hexagonal structure with space group $\mathrm{P6_3/mmc}$. The energies of formation are fitted with the solid solution model of the form,

\begin{equation}\label{eq:solid-sol}
\Delta E(x) = \Omega x(1-x)
\end{equation}

Here, $\Omega$ is the interaction parameter between the nearest-neighbor anion site. The Gibbs free energy is determined by adding $-T\Delta S_{config}$ to $\Delta E(x)$ as given in Equation \ref{eq:Gibbs}.

\begin{equation}\label{eq:Gibbs}
\Delta G(x) = \Delta E(x) -T\Delta S_{config}
\end{equation}

Here, $\Delta S_{config}$ is the configurational entropy of the alloy of composition $x$ and is given by,

\begin{equation}\label{eq:configurational_entropy}
\Delta S_{config} = -k_B[(1-x)ln(1-x) + xlnx]
\end{equation}

The equilibrium concentration of the alloy at any temperature $T$ is obtained from the common tangent to the Gibbs free energy curves.

The ground states of the alloy in the hexagonal lattice are determined with the aid of the CE method \citep{Ducastelle1991}. Drawn from the Ising model for magnetic materials, the CE formalism expresses the energy of formation of an alloy $\mathrm{A_{1-x}B_{x}}$ with configuration $\tau$, as the sum of the effective cluster interaction, $\xi_{i}$ multiplied with the average correlation, $\overline{\Xi}_{i}(\tau)$  associated with a cluster of lattice sites, $i$ given by,

\begin{equation}\label{eq:ce}
\Delta E(\tau) = N\sum_{i}^{I}J_{i}\overline{\Xi}_{i}(\tau)\xi_{i} 
\end{equation}

Here, $N$ is the total number of lattice sites, $J_{i}$ is the number of equivalent clusters $i$ and I is the largest cluster where CE is truncated. The variable $\tau$, represents the different possible arrangements of A and B at the lattice sites in $\mathrm{A_{1-x}B_{x}}$. Each such arrangement is called a configuration of the alloy. $\xi_{i}$ can be obtained from Equation \ref{eq:ce} by fitting $\Delta E(\tau)$ calculated by DFT and $N$, $J_{i}$ and $i$ as given in Equation \ref{eq:eci} \citep{Wei1990}.

\begin{equation}\label{eq:eci}
\xi_{i} = \frac{1}{NJ_{i}}\sum_{\tau}^{T}[\overline{\Xi}_{i}]^{-1}\Delta E(\tau) 
\end{equation}

The CE for the energy of formation of $\mathrm{KCuTe_{1-x}Se_{x}}$ in the P$6_3$/mmc structure is generated using the Automated Alloy Theoretic Toolkit (ATAT) by fitting 55 ordered structures \citep{avdw:atat, Walle_2002, avdw:atat2}. The force constants of all the ordered structures are determined from the bond length-dependent transferable force constants which are further used to determine the phonon frequencies and hence, the phonon density of states of the structures. Force constants are calculated for a small subset of the structures that are used to obtain the CE of the energy of formation. The size of the supercells of these structures are such that they inscribe a sphere of radius 10\AA. 1\% strain is introduced in the supercells with 4 intermediate strain levels. The atoms are perturbed from their equilibrium positions by 0.2\AA. The dependence of the force constant on the bond length is obtained by a linear fit. This linear equation is then used to determine the force constants of all the ordered structures which are further used to determine the phonon frequencies and the phonon density of states. The vibrational free energy of the structures at different temperatures are obtained by integrating the phonon density of states. A CE for the vibrational free energy is generated which is added to the CE of the energy of formation.

The temperature-composition phase diagram of the alloy is also calculated via CE-based MC simulations in the semi-grand canonical ensemble. In this ensemble, the $\mathrm{A_{1-x}B_{x}}$ system is considered to be in contact with a heat and chemical reservoir. This ensures that the temperature, $\beta = \frac{1}{k_{B}T}$ and chemical potential, $\mu = \mu_{A}-\mu_{B}$, of the system are constant. However, the energy and number of atoms of A and B can fluctuate to maintain equilibrium. The partition function of a semi-grand canonical ensemble is given by Equation \ref{eq:partition_fn}

\begin{equation}\label{eq:partition_fn}
Z = \sum_{\tau}e^{-\beta N[\Delta E(\tau)-\mu x(\tau)]}
\end{equation}

The semi-grandcanonical potential per atom can be obtained from the partition function with Equation \ref{eq:phi1},

\begin{equation}\label{eq:phi1}
\beta \phi = -\frac{1}{N}lnZ
\end{equation}

Taking derivative of Equation \ref{eq:phi1} and using Equation \ref{eq:partition_fn} gives,

\begin{equation}\label{eq:phi2}
d(\beta \phi) = (\Delta E - \mu x)d\beta - \beta x d\mu
\end{equation}

Integrating Equation \ref{eq:phi2} gives the semi-grand canonical potential, $\phi$ at any $(\beta_{1},\mu_{1})$.

\begin{equation}\label{eq:phi3}
\beta_{1}\phi(\beta_{1},\mu_{1}) = \beta_{0}\phi(\beta_{0},\mu_{0}) + \int_{\beta_{0},\mu_{0}}^{\beta_{1},\mu_{1}}[(\Delta E - \mu x)d\beta - \beta x d\mu]
\end{equation} 

The temperature-composition phase diagram of the alloy system $\mathrm{A_{1-x}B_{x}}$ is obtained by tracing the intersection of the potential surfaces $\phi^{A}(\beta,\mu)$ and $\phi^{B}(\beta,\mu)$ and the equilibrium concentration of the alloy at temperature $\beta$ is given by the derivative of $\phi^{A/B}$ with respect to $\mu$ at the point of intersection. The phase boundary of $\mathrm{KCuTe_{1-x}Se_{x}}$ is traced by MC simulations on a simulation cell size of  43$\times$43$\times$17 unit cells. MC steps are performed until the average energy has a precision of 0.001 eV/atom.

The electronic bandstructure is calculated by performing a self-consistent field (SCF) calculation with 12800 KPPRA followed by a non self-consistent field calculation along the high symmetry k path of $\mathrm{\Gamma}$ - X $\vert$ Y - $\mathrm{\Gamma}$ - Z $\vert$ R - $\mathrm{\Gamma}$ -T $\vert$ U - $\mathrm{\Gamma}$ - V in the Brillouin zone. The bandgaps of  $\mathrm{KCuTe_{1-x}Se_{x}}$ structures are determined using the PBE as well as the meta-GGA functional, mBJ \citep{Tran2009}. The optical properties of the alloy are determined from the dielectric constant, calculated with PBE under the independent particle approximation (IPA) with twice the number of bands used in the electronic bandstructure calculation \citep{Adler1962, Wiser1963}. 

The alignment of the conduction band edge and valence band edge with the water redox potential is determined by the Butler-Ginley model given by Equation \ref{eq:butler-ginley} \citep{Butler_1978}.

\begin{equation}\label{eq:butler-ginley}
    E_{V/C} = E_{0} + (\chi_{K} \chi_{Cu} \chi_{Te}^{(1-x)} \chi_{Se}^{x})^{\frac{1}{3}} \pm \frac{E_{g}}{2}
\end{equation}

Here, $E_{0}$ is the potential of the normal hydrogen electrode (NHE) with respect to the vacuum and is equal to -4.5 eV. $\chi_{K}$, $\chi_{Cu}$, $\chi_{Te}$ and $\chi_{Se}$ are the electronegativities of elements K, Cu, Te and Se, respectively, on the Mulliken scale and $E_{g}$ is the mBJ-calculated band gap of the alloy.

\section{Results and Discussion}
KCuTe has been reported theoretically and experimentally in the $\mathrm{P6_3/mmc}$ structure, while KCuSe has been reported in the $\mathrm{P6_3/mmc}$, $\mathrm{Pna2_1}$ and Pnma structures, respectively. The energies of the candidate structures for KCuTe and KCuSe with respect to the energy of the $\mathrm{P6_3/mmc}$ structure are given in Table \ref{stability}. While the lowest energy structure for KCuTe is the hexagonal $\mathrm{P6_3/mmc}$ structure, the orthorhombic Pnma structure is the lowest energy for KCuSe. Another point to note is that the positions of all ions, as reported in experiments \citep{Savelsberg1978}, in the hexagonal structures, are agnostic to the anion, whereas that is not the case for the orthorhombic structure. Although KCuSe has the same spacegroup, the orthorhombic structures of KCuTe have different crystal coordinates for K, Cu, and Te compared to the crystal coordinates of K, Cu, and Se in KCuSe. This is extremely important because we are interested in the substitutional alloys of KCuTe and KCuSe.  As a result, the alloy system $\mathrm{KCuTe_{1-x}Se_x}$ cannot be studied in the orthorhombic structure within the SQS and CE framework. We have chosen the experimentally reported hexagonal structure to study $\mathrm{KCuTe_{1-x}Se_x}$ alloy.

	\begin{table}[!htbp]
        \centering
	\caption{Relative energy of KCuTe and KCuSe in P$6_3$/mmc, Pna$2_{1}$ and Pnma structures with respect to the P${6_3}$/mmc structure.}
	\begin{tabular}{ ccc}
		\hline
		Structure & KCuTe (eV/f.u.) & KCuSe (eV/f.u.) \\
		\hline
		$\mathrm{P6_3/mmc (\alpha)}$ & 0 & 0 \\
		$\mathrm{Pna2_1 (\beta)}$ & 0.2417 & -0.0028 \\
		$\mathrm{Pnma (\gamma)}$ & 0.1813 & -0.0205 \\
		\hline
	\end{tabular}
	\label{stability}
	\end{table}

\subsection{Phase diagram using solid solution model}

\begin{figure}[!hbtp]
\centering
\includegraphics[width=0.45\textwidth]{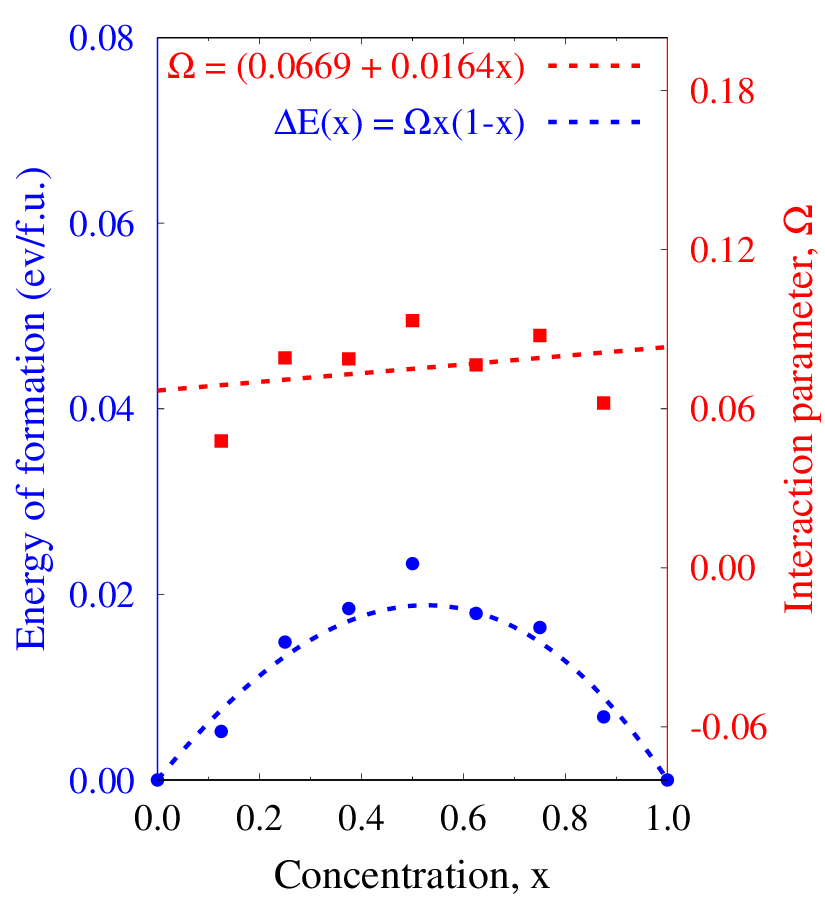}
\caption{Plot of the energy of formation of SQS of $\mathrm{KCuTe_{1-x}Se_{x}}$ in the hexagonal structure. The blue axis corresponds to the energy of formation and the red axis corresponds to the interaction parameter, $\Omega$.}
\label{fig:kcutese_solid_solution}
\end{figure}

The energy of formation of the SQS of $\mathrm{KCuTe_{1-x}Se_{x}}$ at different compositions is shown in blue in Figure \ref{fig:kcutese_solid_solution}. Unlike an ideal solid solution, the distribution of the formation energy is not symmetric about $x$ = 0.5 and is skewed to the right. To gain a deeper understanding, we have calculated the interaction parameter, $\Omega$, corresponding to each SQS. $\Omega$ increases linearly with Se concentration as shown in red in Figure \ref{fig:kcutese_solid_solution}. Consider the alloy with complementary compositions $x_{1}$ and $x_{2}$ such that $x_{2}$ = 1 - $x_{1}$ and $x_{2} \ge x_{1}$. In an ideal solid solution, $\Delta E(x_{1})$ should have been equal to $\Delta E(x_{2})$. However, since $\Omega_{x_{1}} \le \Omega_{x_{2}}$, therefore, $\Delta E(x_1) \le \Delta E(x_2)$. The energy of formation of the alloy at the Se-rich end is greater than that at the Te-rich end, implying that it is easier to incorporate the smaller anion (Se) in the larger anion (Te) sublattice as compared to incorporating the larger anion in the smaller anion sublattice. This also indicates that because of the lattice mismatch between KCuTe and KCuSe, the interaction between lattice sites beyond the nearest--neighbor pair contributes to the energetics of the alloy.

	\begin{table*}[!hbtp]
      \centering
	\caption{The contribution of volume deformation (VD), charge exchange (ChE) and structure relaxation (SR) towards the energy of formation of $\mathrm{KCuTe_{1-x}Se_{x}}$ at different concentrations in the hexagonal structure. }
	\begin{tabular}{ ccccc}
	\hline
	x & $\mathrm{\Delta E^{total}}$(eV/f.u.) & $\mathrm{\Delta E^{VD}}$(eV/f.u.) & $\mathrm{\Delta E^{ChE}}$(eV/f.u.) & $\mathrm{\Delta E^{SR}}$(eV/f.u.) \\
	\hline
	0.125 & 0.0052 & 0.1079 & -0.0091 & -0.0936 \\
	0.25 & 0.0149 & 0.0971 & 0.0139 & -0.0962  \\
	0.375 & 0.0185 & 0.0803 & -0.0017 & -0.0601  \\
	0.5 & 0.0233 & 0.0806 & -0.0127 & -0.0445 \\
	0.625 & 0.0179 & 0.0879 & -0.0084 & -0.0616  \\
	0.75 & 0.0164 & 0.1333 & -0.0168 & -0.1001 \\
	0.875 & 0.0068 & 0.1339 & -0.0195 & -0.1076 \\
	\hline
	\end{tabular}
	\label{table:VD_CE_SR_eof_kcutese}
	\end{table*}

The energetics of the alloy with composition $x$ is governed by three major physical contributions, namely volume deformation (VD), charge exchange (ChE) and structure relaxation (SR) \citep{Srivastava1985, Ferreira1989,Sanati2003}. VD arises due to the contraction and expansion of KCuTe and KCuSe, respectively, to the lattice constant of the alloy $a(x)$. When these volume--deformed compounds are mixed to form a solid solution, the electronic interaction between the Te and Se anions results in the ChE term. Finally, the change in energy arising from relaxation of the atomic positions contributes to the SR term. This decomposition of the formation energy is given in Table \ref{table:VD_CE_SR_eof_kcutese}. As expected, $\Delta E^{VD} >$ 0 in $\mathrm{KCuTe_{1-x}Se_{x}}$ due to lattice mismatch. Although SR lowers the energy, $\Delta E^{VD}+ \Delta E^{SR}$ remains positive. The energy associated with ChE $<$ 0 and the strength of $\Delta E^{ChE}$ increases as the lattice constant decreases with increasing Se concentration. However, the chemical interaction between Te and Se is not strong enough to overcome the positive contribution from the lattice distortion, resulting in net positive energy of formation at all compositions. This indicates the presence of a miscibility gap in the alloy.

\begin{figure*}[!hbtp]
\centering
\includegraphics[width=\textwidth]{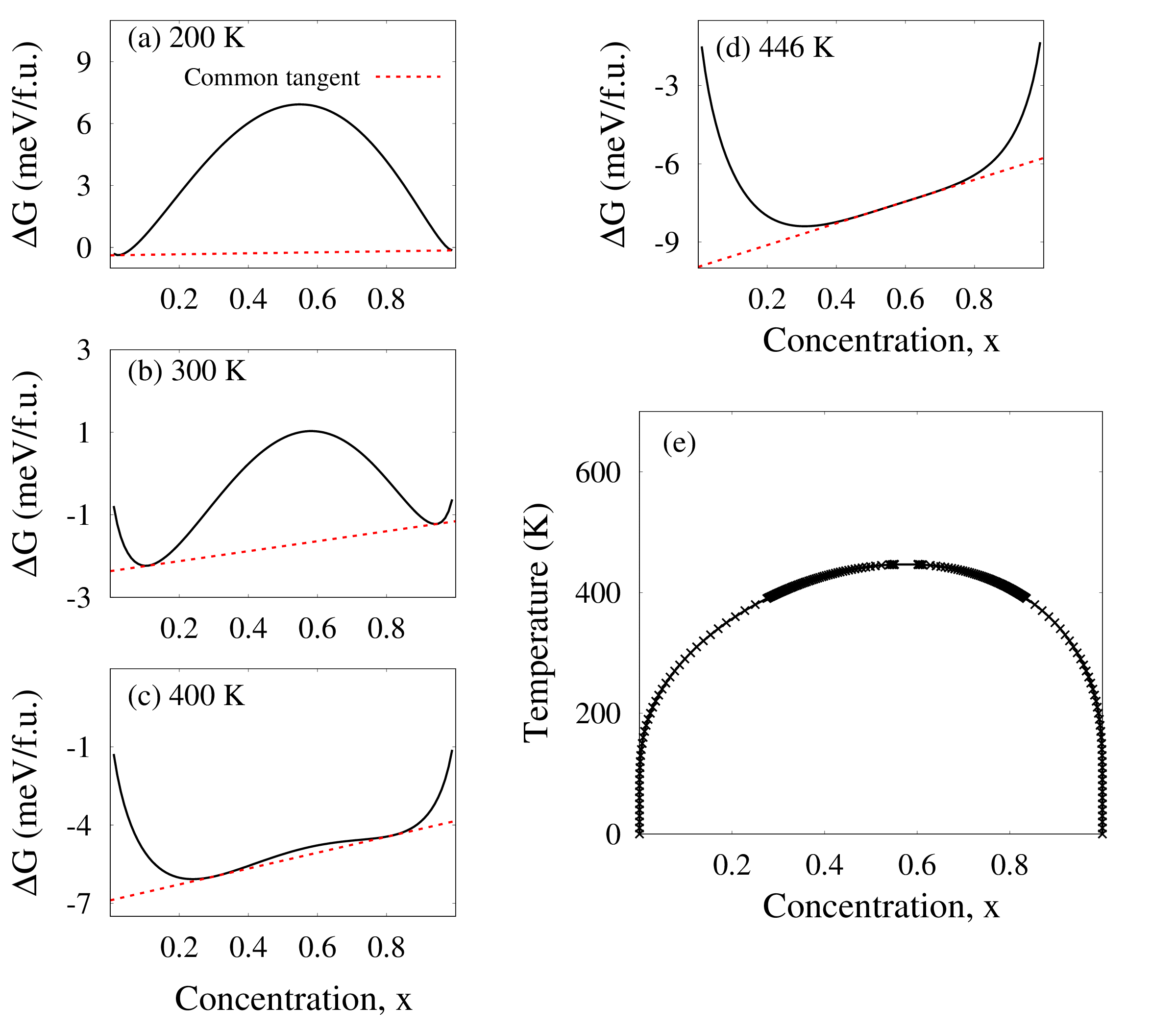}
\caption{(a)-(d) Plots of the variation of Gibb's free energy with the concentration of $\mathrm{KCuTe_{1-x}Se_{x}}$ alloy at different temperatures. (e) Temperature-composition phase diagram calculated from the solid solution model. The black crosses are the equilibrium concentration at different temperatures and the black line is an envelop of the data points.}
\label{fig:kcutese_free_energy}
\end{figure*}

The Gibbs free energy of $\mathrm{KCuTe_{1-x}Se_{x}}$ calculated from equations \ref{eq:Gibbs} and \ref{eq:configurational_entropy} at different temperatures is shown in Figure \ref{fig:kcutese_free_energy}. At this point, we are not considering any other phases to exist over the entire composition range; only the hexagonal phase is stable. In such a situation, a single \emph{$\Delta$G vs. x} curve is representative of the alloy system, as given by the regular solution model. Consider the \emph{$\Delta$G vs. x} curve at T = 200 K, \ref{fig:kcutese_free_energy}(a). For $0 \leq \mathrm{x} \leq 0.024$, the free energy of the alloy decreases with increasing x. Therefore, for any value of x in this range, the system exists as a single alloy. A similar situation is found for alloys with composition in the range $0.991 \leq \mathrm{x} \leq 1$. However, in the composition range $0.024 \leq \mathrm{x} \leq 0.991$, a single alloy has a higher energy than a mechanical mixture of alloys with composition 0.024 and 0.991, the common tangent construction.  When the temperature increases to 300 K, the configurational entropy contributes significantly to the Gibbs free energy, giving rise to a clearer W-shaped curve, \ref{fig:kcutese_free_energy}(b). The common tangent passes at $x_{1} =$ 0.110 and $x_{2} =$ 0.945. The solubility of Se at the Te-rich and that of Te at the Se-rich increases significantly as the temperature rises to 400 K, \ref{fig:kcutese_free_energy}(d). The equilibrium concentrations of the phase separated alloy are $x_{1} =$ 0.302 and $x_{2} =$ 0.816.  From 446 K onward, the Gibbs free energy is a convex curve. As a result, the alloy does not phase separate and forms a solid solution over the entire composition range.

Putting these observations in a temperature-composition phase diagram (Figure \ref{fig:kcutese_free_energy} (d)), a miscibility gap is the result, with a critical point at T = 446 K and x = 0.601. Within the miscibility gap the alloy tends to phase separate. Beyond the critical temperature of 446 K, the alloy forms a solid solution in the hexagonal structure throughout the composition range. Since the solid solution model does not take into account the possibility of ordering in $\mathrm{KCuTe_{1-x}Se_{x}}$, we move to a more sophisticated CE method to search for the occurrence of ordered compounds.

\subsection{Phase diagram using CE based MC simulations}

\begin{figure*}[!hbtp]
\centering
\includegraphics[width=\textwidth]{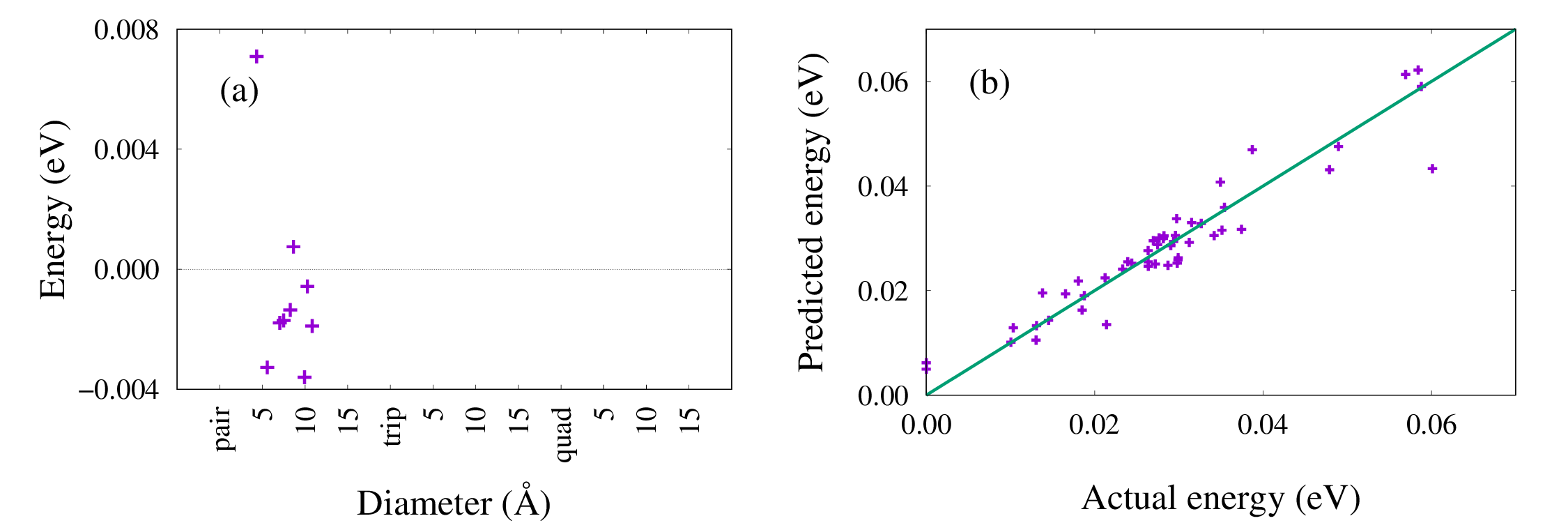}
\caption{Plot of the (a) effective cluster interactions and (b) CE predicted energy versus actual energy of different ordered structures of $\mathrm{KCuTe_{1-x}Se_{x}}$.}
\label{fig:kcutese_CE}
\end{figure*}

Figure \ref{fig:kcutese_CE}(a) is a graphical representation of the interaction energies of the various clusters present in the cluster expansion. Interactions of up to nine nearest-neighbor pairs of lattice sites are sufficient to accurately predict the energy of formation of the ordered structures. The strength of interaction is largest between the nearest--neighbor sites of the anion sublattice. As we move farther away from a lattice site, the strength of the interaction is reduced. The error associated with the CE of $\mathrm{KCuTe_{1-x}Se_{x}}$ is given by the cross-validation score and is equal to 0.0054 eV/f.u. Figure \ref{fig:kcutese_CE}(b) compares the energy of formation of 49 ordered structures predicted by the CE with that calculated from DFT. It reflects the predictive power of CE. The energy of formation is distributed closely around the $y=x$ line for lower energy structures, whereas the energy of formation is scattered away from the line for higher energy structures. Therefore, CE accurately predicts the energy of formation of structures with lower energy. This is adequate for an accurate prediction of the ground states.

\begin{figure}[!htbp]
\centering
\includegraphics[width=0.45\textwidth]{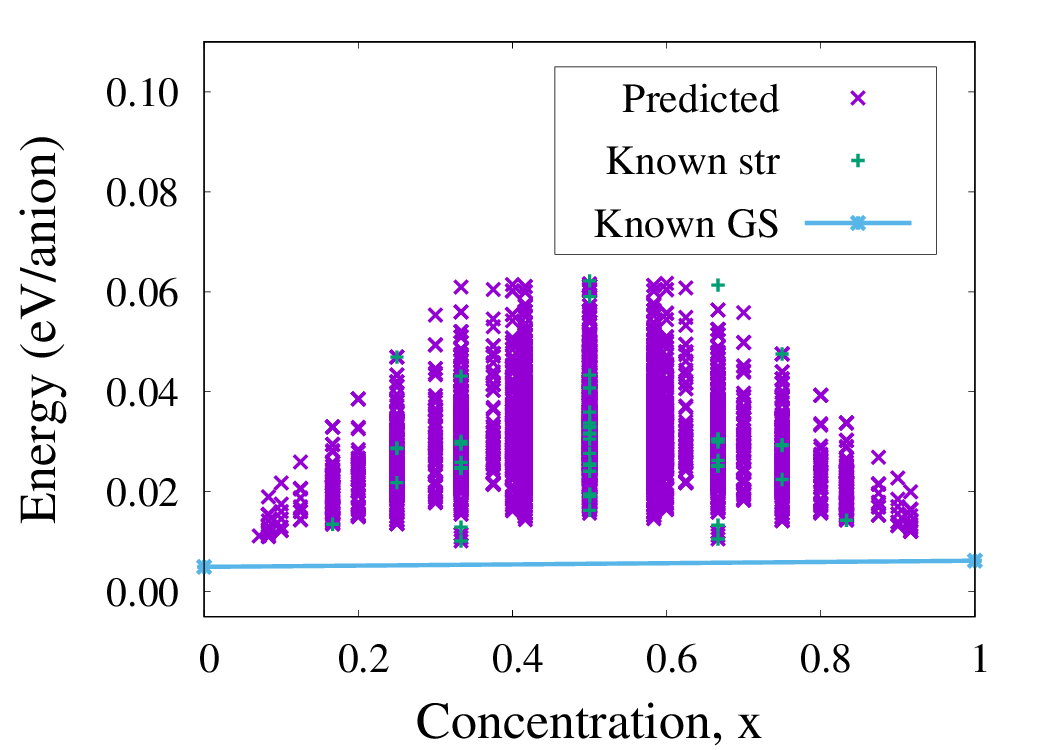}
\caption{Plot of the energy of formation of ordered structures of different concentrations and configurations of anions of $\mathrm{KCuTe_{1-x}Se_{x}}$ calculated from CE.}
\label{fig:kcutese_Fitted_energy}
\end{figure}

The energy of formation of ordered structures of $\mathrm{KCuTe_{1-x}Se_{x}}$ predicted by CE is shown in Figure \ref{fig:kcutese_Fitted_energy}. Since the energies of formation of all the structures are greater than the composition--weighted average energy represented by the blue line, therefore, KCuTe and KCuSe in the hexagonal structure are the only two ground states of the $\mathrm{KCuTe_{1-x}Se_{x}}$ alloy system. This lends credence to the initial assumption regarding the use of only the hexagonal phases for both KCuTe and KCuSe to obtain the phase diagram.

Due to the lattice mismatch between KCuTe and KCuSe, the equilibrium distance between the Cu-K, Cu-Te, Cu-Se, K-K, K-Se, and K-Te atoms in the various ordered structures of $\mathrm{KCuTe_{1-x}Se_{x}}$ is different from the equilibrium distances between these atoms in pure KCuTe and pure KCuSe. Moreover, the distances between the atoms vary with the composition and anion configuration of the ordered structures. As a result, establishing a relation between the force constant versus the bond length associated with stretching and bending aids to calculate the force constant matrix and the phonon frequencies for a large number of structures used to generate the CE. The transferable force constants, dependent on the bond length, of $\mathrm{KCuTe_{1-x}Se_{x}}$ are shown in Figure S1 (supplementary material). The term ``bond" in this discussion does not necessarily imply the sharing of electrons between two atoms A and B. Rather, it implies the force experienced by atom A as a result of a perturbation in the position of B and vice versa. The magnitude of the force constant is significant in the case of Cu-Te and Cu-Se as a result of shorter bond lengths as compared to Cu-K, K-K, K-Te, and K-Se. The magnitude of the force constant associated with stretching decreases with an increase in the distance between Cu and Te and between Cu and Se. However, the force constant associated with bending increases with an increase in the bond length (Figures S1(b) and S1(c)).  For bonds longer than 3 \AA, as in the case of Cu-K, K-K, K-Te, and K-Se, there is negligible variation in the magnitude of the force constant with increasing bond length. The dependence of the force constant on the bond length, modeled using a linear equation, was used to determine the force constants and the phonon frequencies, which were in turn used to calculate the vibrational free energy of the ordered structures used to fit the CE for the energy of formation. Since the vibrational free energy also depends on the configuration $\sigma$,  a CE of the vibrational free energy at different temperatures is added to the CE of the formation energy.

\begin{figure}[!htbp]
\centering
\includegraphics[width=0.45\textwidth]{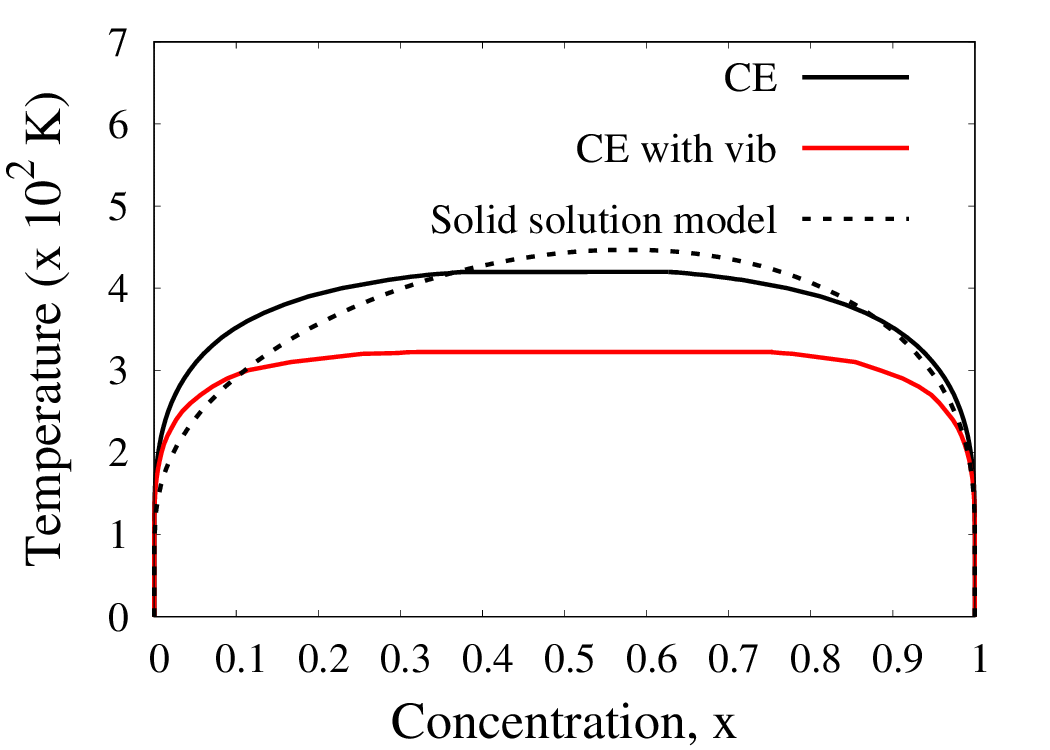}
\caption{The T-x phase diagram of $\mathrm{KCuTe_{1-x}Se_{x}}$ calculated from CE based MC simulations with and without the contribution of vibrational free energy compared with that obtained from the solid solution method. }
\label{fig:kcutese_combined_pd}
\end{figure}

The temperature-composition phase diagram of $\mathrm{KCuTe_{1-x}Se_{x}}$ calculated from CE--based MC simulations is shown in Figure \ref{fig:kcutese_combined_pd}. The phase diagram predicts a miscibility gap up to a peak critical temperature of 420 K, below which the alloy phase is separated into two solid solutions of different compositions. Beyond 420 K, the alloy forms a uniform solid solution in the hexagonal structure over the entire composition range. When the effect of lattice vibrations towards the thermodynamic stability of the alloy is taken into account, the peak critical temperature reduces from 420 K to 322 K. As expected, lattice vibrations enhance the solubility of Se at the Te-rich end and that of Te at the Se-rich end. The critical temperature of 420 K predicted by the CE-based MC simulations is lower than that predicted by the solid solution model (446 K). Because CE-based MC simulations not only take into account long--range interactions of the anion sublattice but also perform a statistical sum of the thermodynamic quantities, therefore they are more accurate as compared to the solid solution model. However, both models predict the formation of a hexagonal solid solution over the entire composition range above the critical temperature.

\subsection{Electronic bandstructure}

The bandgap of $\mathrm{KCuTe_{1-x}Se_{x}}$ at different compositions calculated using PBE and mBJ are given in Table \ref{kcutese_electronic_structure_properties}.

	\begin{table*}[hpbt!]
	\caption{Bandgap and effective mass of charge carriers of $\mathrm{KCuTe_{1-x}Se_{x}}$ at different concentrations. In the column containing the effective mass of the charge carriers, the first row at each concentration is the effective mass of holes, $m_{h}^{*}/m_{0}$, at the valence band maxima (VBM) and the second row is the effective mass of electrons, $m_{e}^{*}/m_{0}$ at the conduction band minima (CBM). }
	\begin{tabular}{ cccccccccc}
		\hline
		\multirow{2}{*}{\textbf{x}} & \multicolumn{2}{c}{\textbf{$\mathrm{E_g}$ (eV)}} & \multicolumn{7}{c}{Effective mass} \\
		& PBE & mBJ & \multicolumn{2}{c}{\textbf{$\mathrm{\Gamma-M}$}} & \multicolumn{3}{c}{\textbf{$\mathrm{K-\Gamma}$}} & \multicolumn{2}{c}{\textbf{$\mathrm{\Gamma-Z}$}} \\
		\hline
		\multirow{2}{*}{0} & \multirow{2}{*}{0.611} & \multirow{2}{*}{1.466} & \multicolumn{2}{c}{-0.462} & \multicolumn{3}{c}{-0.455} & \multicolumn{2}{c}{-4.54} \\
		& & & \multicolumn{2}{c}{0.153} & \multicolumn{3}{c}{0.178} & \multicolumn{2}{c}{0.412} \\ 
		& & & & & & & & & \\
		\multirow{2}{*}{1} & \multirow{2}{*}{0.207} & \multirow{2}{*}{1.154} & \multicolumn{2}{c}{-0.541} & \multicolumn{3}{c}{-0.534} & \multicolumn{2}{c}{-9.987} \\
		& & & \multicolumn{2}{c}{0.231} & \multicolumn{3}{c}{0.295} & \multicolumn{2}{c}{0.434} \\ 
		& & & & & & & & & \\	
		\cline{4-10}
		& & & \textbf{$\mathrm{\Gamma-X}$} & \textbf{$\mathrm{Y-\Gamma}$} & \textbf{$\mathrm{\Gamma-Z}$} & \textbf{$\mathrm{R-\Gamma}$} & \textbf{$\mathrm{\Gamma-T}$} & \textbf{$\mathrm{U-\Gamma}$} & \textbf{$\mathrm{\Gamma-V}$} \\
		\cline{4-10}	
		& & & & & & & & & \\	
		\multirow{2}{*}{0.125} & \multirow{2}{*}{0.543} & \multirow{2}{*}{1.420} & -8.117 & -0.528 & -6.374 & -1.169 & -0.935 & -6.229 & -0.672 \\
		& & & 0.427 & 0.008 & 0.007 & 0.127 & 0.007 & 0.123 & 0.009 \\ 
		& & & & & & & & & \\
		\multirow{2}{*}{0.25} & \multirow{2}{*}{0.474} & \multirow{2}{*}{1.370} & -0.535 & -0.535 & -0.007 & -0.023 & -0.511 & -0.682 & -0.493 \\
		& & & 0.083 & 0.083 & 0.068 & 0.016 & 0.081 & 0.081 & 0.011 \\ 
		& & & & & & & & & \\
		\multirow{2}{*}{0.5} & \multirow{2}{*}{0.257} & \multirow{2}{*}{1.230} & -0.060 & -9.316 & -0.496 & -2.756 & -1.425 & -2.108 & -0.915 \\
		& & & 0.077 & 0.466 & 0.006 & 0.124 & 0.156 & 0.086 & 0.122 \\ 
		& & & & & & & & & \\
		\multirow{2}{*}{0.75} & \multirow{2}{*}{0.258} & \multirow{2}{*}{1.190} & -0.653 & -0.654 & -0.011 & -0.314 & -1.51 & -0.577 & -0.554 \\
		& & & 0.069 & 0.07 & 0.048 & 0.122 & 0.069 & 0.069 & 0.115 \\ 
		& & & & & & & & & \\
		\multirow{2}{*}{0.875} & \multirow{2}{*}{0.201} & \multirow{2}{*}{1.140} & -8.544 & -0.895 & -inf & - & - & -inf & -1.019 \\
		& & & 0.427 & 0.005 & 0.003 & 0.007 & 0.004 & 0.006 & 0.007 \\ 
		\hline
	\end{tabular}
	\label{kcutese_electronic_structure_properties}
	\end{table*}

Despite a sufficiently accurate prediction of the topology of the band structure, there is a significant underestimation of the bandgap by PBE. There is an improvement in the bandgap value by mBJ. Although there are no experimentally reported values of the bandgap for these compounds, the bandgap values predicted from our calculations are in close agreement with those reported in other theoretical studies \citep{Parveen2018, Behera2024}. 
 
\begin{figure}[!htbp]
\centering
\includegraphics[width=0.45\textwidth]{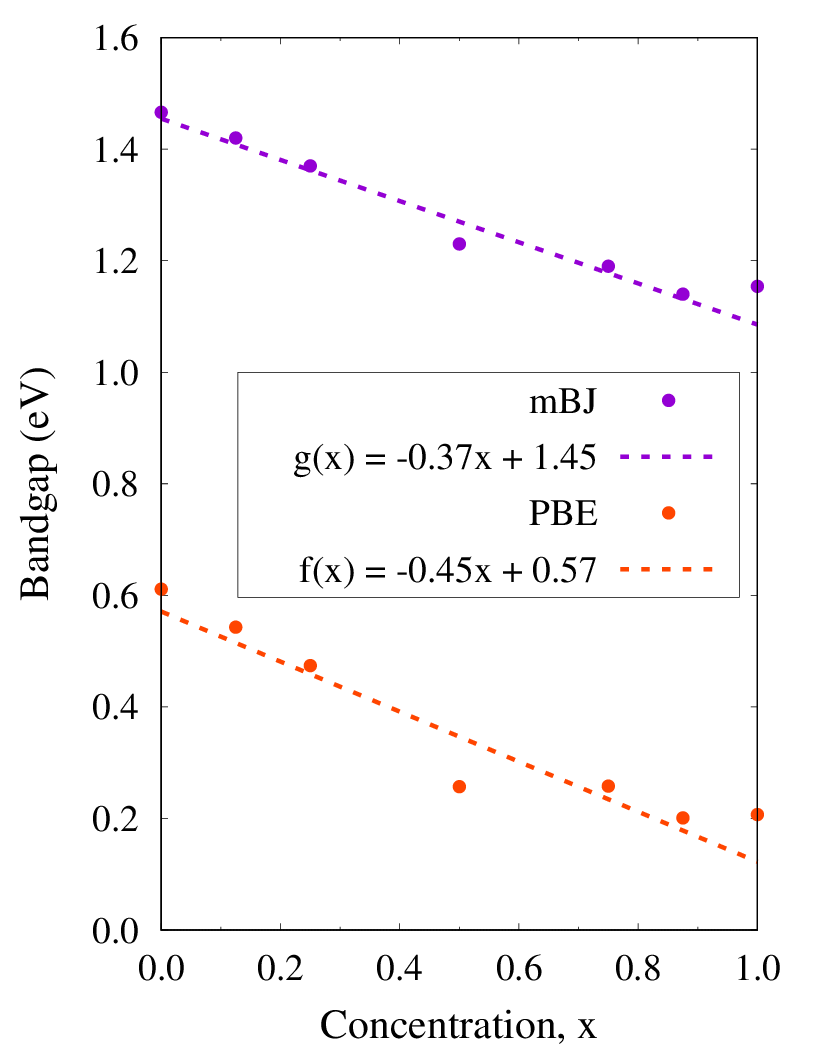}
\caption{Plot of the variation in bandgap with the concentration of Se in $\mathrm{KCuTe_{1-x}Se_{x}}$ calculated with PBE and mBJ exchange-correlation functional.}
\label{fig:kcutese_bandgap_vs_x}
\end{figure}

\begin{figure}[!htbp]
\centering
\includegraphics[width=0.45\textwidth]{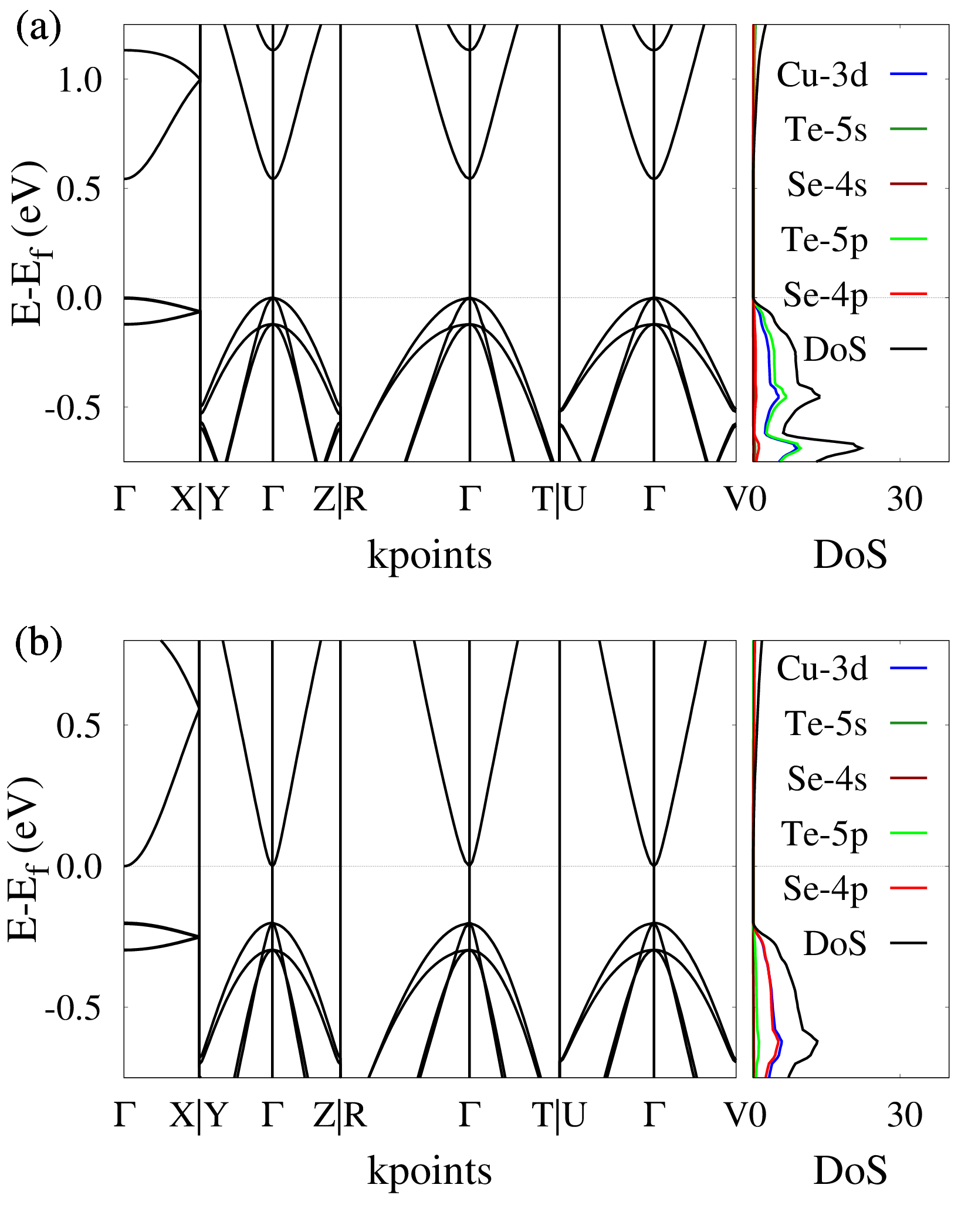}
\caption{Plots of electronic band structure and density of states of $\mathrm{KCuTe_{1-x}Se_{x}}$ at a) $x$ = 0.125 b) $x$ = 0.875.}
\label{fig:kcutese_bs}
\end{figure} 
 
Figure \ref{fig:kcutese_bandgap_vs_x} is a plot of the variation of the bandgap with the alloy concentration. The bandgap of the alloy decreases with an increase in Se concentration. Unlike the bandgap bowing observed in highly mismatched alloys due to band anticrossing, $\mathrm{KCuTe_{1-x}Se_{x}}$ shows Vegard's law behavior of the bandgap. The electronic band structure of $\mathrm{KCuTe_{1-x}Se_{x}}$ at different concentrations calculated with PBE are shown in Figures \ref{fig:kcutese_bs} and S2. KCuTe, KCuSe, and their alloys are direct bandgap semiconductors. The valence band maxima (VBM) and the conduction band minima (CBM) occur at the $\mathrm{\Gamma}$ point. The effective masses of electrons and holes at the band extrema are determined from the band structure using $\frac{1}{m^{*}} = \frac{1}{\hbar^2}\frac{d^{2}E}{dk^{2}}$ and are also given in Table \ref{kcutese_electronic_structure_properties}. The double derivative of $E$ with respect to $k$ was determined by the finite difference method. The effective mass of electrons in the alloy are lower than those in the end compounds. As a result, electrons have higher mobility in the alloys compared to those in the pure compounds, thus enhancing their transport through most of the semiconductor alloy.

	\begin{table}[hpbt!]
	\centering
	\caption{The average energy of Te-5p orbital with respect to Se-4p orbital in the valence band and Se-4s orbital with respect to Te-5s orbital in the conduction band at different compositions of $\mathrm{KCuTe_{1-x}Se_{x}}$ in hexgonal structure. }
	\begin{tabular}{ ccc}
	\hline
	x & $\mathrm{Te_{5p} - Se_{4p}}$(eV) & $\mathrm{Se_{4s} - Te_{5s}}$(eV) \\
	\hline
	0.125 & 0.341 & -0.312 \\
	0.25 & 0.247 & -0.321   \\
	0.5 & -0.38 & 0.075  \\
	0.75 & 0.255 & -0.329  \\
	0.875 & 0.33 & -0.337 \\
	\hline
	\end{tabular}
	\label{table:bandcrossing_kcutese}
	\end{table}

The conduction band of the alloy is primarily composed of Se-4s and Te-5s states, while the valence band is composed of Cu-3d, Se-4p and Te-5p states (Figure \ref{fig:kcutese_bs}). The concentrations of the alloy that have been taken into consideration in this study are beyond the impurity regime. As a result, Te-5p states at the Se-rich end and Se-4s states at the Te-rich end form extended bands within the valence band and conduction band, respectively. The dominant states at the VBM change from Se-4p to Te-5p with increasing Te concentration. Similarly, the characteristics of the CBM change from Te-5s to Se-4s with increasing Se concentration. The center of Se-4p and Te-5p in the valence band and Se-4s and Te-5s in the conduction band are the weighted average energy. In VB, the band center of Te-5p is always higher in energy than Se-4p, and in CB, the band center of Se-4s is lower in energy than Te-5s (see Table \ref{table:bandcrossing_kcutese}). This is an essential property of the electronic band structure of HMAs such as $\mathrm{GaAs_{1-x}Bi_{x}}$ and $\mathrm{GaAs_{1-x}N_{x}}$ \citep{Deng2010}. Despite satisfying the essential band structure criterion of HMAs, linear variation in the bandgap is counterintuitive. The decomposition of the bandgap into the contributions of VD, ChE, and SR, given in Table \ref{table:VD_CE_SR_bandgap_kcutese} will provide a deeper insight into the system.

	\begin{table}[!hpbt]
	\caption{The contribution of volume deformation (VD), charge exchange (ChE) and structure relaxation (SR) towards the bandgap of $\mathrm{KCuTe_{1-x}Se_{x}}$ at different concentrations in the hexagonal structure calculated with PBE exchange-correlation functional. }
	\begin{tabular}{ cccccc}
	\hline
	x & $\mathrm{E^{Vegard}_{g}}$(eV) & $\mathrm{\Delta E^{PBE-Vegard}_{g}}$(eV) & $\mathrm{E^{VD}_{g}}$(eV) & $\mathrm{E^{ChE}_{g}}$(eV) & $\mathrm{E^{SR}_{g}}$(eV) \\
	\hline
	0 & 0.611 & 0 & 0 & 0 & 0 \\
	0.125 & 0.561 & -0.0175 & -0.0745 & -0.0434 & 0.1004 \\
	0.25 & 0.510 & -0.036 & -0.0226 & -0.0561 & 0.0427  \\
	0.5 & 0.409 & -0.152 & -0.0230 & -0.0960 & -0.0331 \\
	0.75 & 0.308 & -0.05 & -0.0496 & -0.0076 & 0.0072 \\
	0.875 & 0.258 & -0.0565 & -0.0890 & 0.0142 & 0.0183 \\
	1 & 0.207 & 0 & 0 & 0 & 0 \\
	\hline
	\end{tabular}
	\label{table:VD_CE_SR_bandgap_kcutese}
	\end{table}

Despite the underestimation in the bandgap, PBE exchange-correlation functional effectively captures the trends in the electronic bandgap and the bandstructure. Therefore, we have studied the decomposition of the PBE calculated bandgaps \citep{Wei1995}. The overall bandgap of the alloy at different compositions is less than the average bandgap obtained from Vegard's law, $E_{g}^{Vegard}$. In Table \ref{table:VD_CE_SR_bandgap_kcutese}, $\Delta E_{g}^{PBE-Vegard}$ represents the difference between $E_{g}^{PBE}$ and $E_{g}^{Vegard}$. The contributions of VD, ChE and SR have also been reported with respect to $E_{g}^{Vegard}$. VD and ChE decrease the bandgap from the average value, whereas SR competes with the effect of VD and ChE and increases the bandgap.

\begin{figure}[!htbp]
\centering
\includegraphics[width=0.45\textwidth]{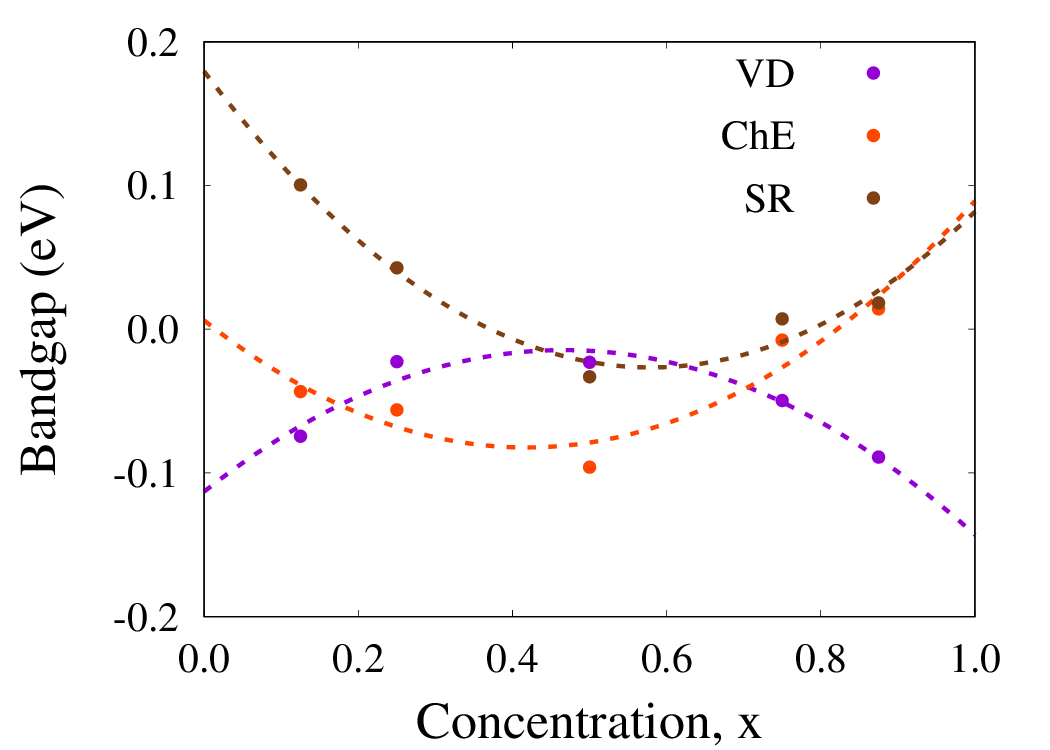}
\caption{Plot of the variation in volume deformation (VD), charge exchange (ChE) and structure relaxation (SR) decomposition of bandgap with the concentration of Se in $\mathrm{KCuTe_{1-x}Se_{x}}$.}
\label{fig:VD_CE_SR_bandgap_kcutese}
\end{figure}

$E_{g}^{ChE}$ and $E_{g}^{SR}$ versus $x$ show a bowing behavior typical of highly mismatched semiconductor alloys (Figure \ref{fig:VD_CE_SR_bandgap_kcutese}). However, $E_{g}^{VD}$ versus $x$ shows a reverse bowing trend. $E_{g}^{VD}(x) \le$ 0 at all alloy concentrations. $E_{g}^{SR}(x) \le$ 0 when 0.37 $\le x \le$ 0.79; it is $\ge$ 0. However, $E_{g}^{ChE}(x) \ge$ 0 only when $x \ge$ 0.82. As a result, SR competes with VD + ChE when $x <$ 0.37. SR favors the lowering of the bandgap when 0.37 $\le x \le$ 0.79. Beyond $x =$ 0.82, ChE + SR compete together with VD. The overall effect of VD + ChE + SR produces a linear variation in bandgap with $x$.

\subsection{Optical Properties}
In this section we discuss the optical absorption properties of the alloy at different Se concentrations. We develop insight into the evolution of the optical absorption spectra with alloy concentration through the real and imaginary parts of the dielectric function, $\varepsilon(\omega) = \varepsilon_{1}(\omega) + i\varepsilon_{2}(\omega)$ computed under the independent electron approximation from the electronic band structure calculated with PBE. Here $\omega$ is the frequency of the electromagnetic wave. The imaginary part of the dielectric constant is obtained from 

\begin{eqnarray}\label{eq:im_epsilon}
    \varepsilon_{2,\alpha \beta}(\omega) = &&\frac{4\pi^{2}e^2}{m^{2}\omega^{2}}\sum_{i,f} \int \bigg[\bra{f}p_{\alpha}\ket{i}\bra{i}p_{\beta}\ket{f} \nonumber \\
    &&\times W_{i}(1-W_{f})\delta(E_{f}-E_{i}-\hbar\omega)\bigg]d^{3}k
\end{eqnarray}

Here, $\bra{f}p_{\alpha}\ket{i}$ and $\bra{i}p_{\beta}\ket{f}$ are the dipole matrix elements corresponding to the crystal directions $\alpha$ and $\beta$. $f, i$ are the final and initial states of transition, respectively. $W_{n}$ is the probability of occupation of the $n$th state determined from the Fermi-Dirac statistics \citep{Bechstedt2006}. 

The real part of the dielectric constant is computed using the Kramers-Kronig relation given by, 

\begin{equation}\label{eq:re_epsilon}
    \varepsilon_{1,\alpha \alpha}(\omega) = 1 + \frac{1}{\pi} P \int_{0}^{\infty}\frac{\omega'\varepsilon_{2}(\omega')_{\alpha \alpha}}{\omega'^{2} - \omega^{2}}d\omega'
\end{equation}

Here, P is the principal value of the integral \citep{ashcroft1976solid}. The real and imaginary parts of the refractive index are related to the dielectric constant by Equations \ref{eq:re_refractive index} and \ref{eq:im_refractive index}, respectively.

\begin{equation}\label{eq:re_refractive index}
    n_1 = \Bigg(\frac{\sqrt{\varepsilon_{1}^{2} + \varepsilon_{2}^{2}} + \varepsilon_{1}}{2}\Bigg)^{\frac{1}{2}}
\end{equation}
\begin{equation}\label{eq:im_refractive index}
    n_2 = \Bigg(\frac{\sqrt{\varepsilon_{1}^{2} + \varepsilon_{2}^{2}} - \varepsilon_{1}}{2}\Bigg)^{\frac{1}{2}}
\end{equation}

\begin{figure*}[!htbp]
\centering
\includegraphics[width=\textwidth]{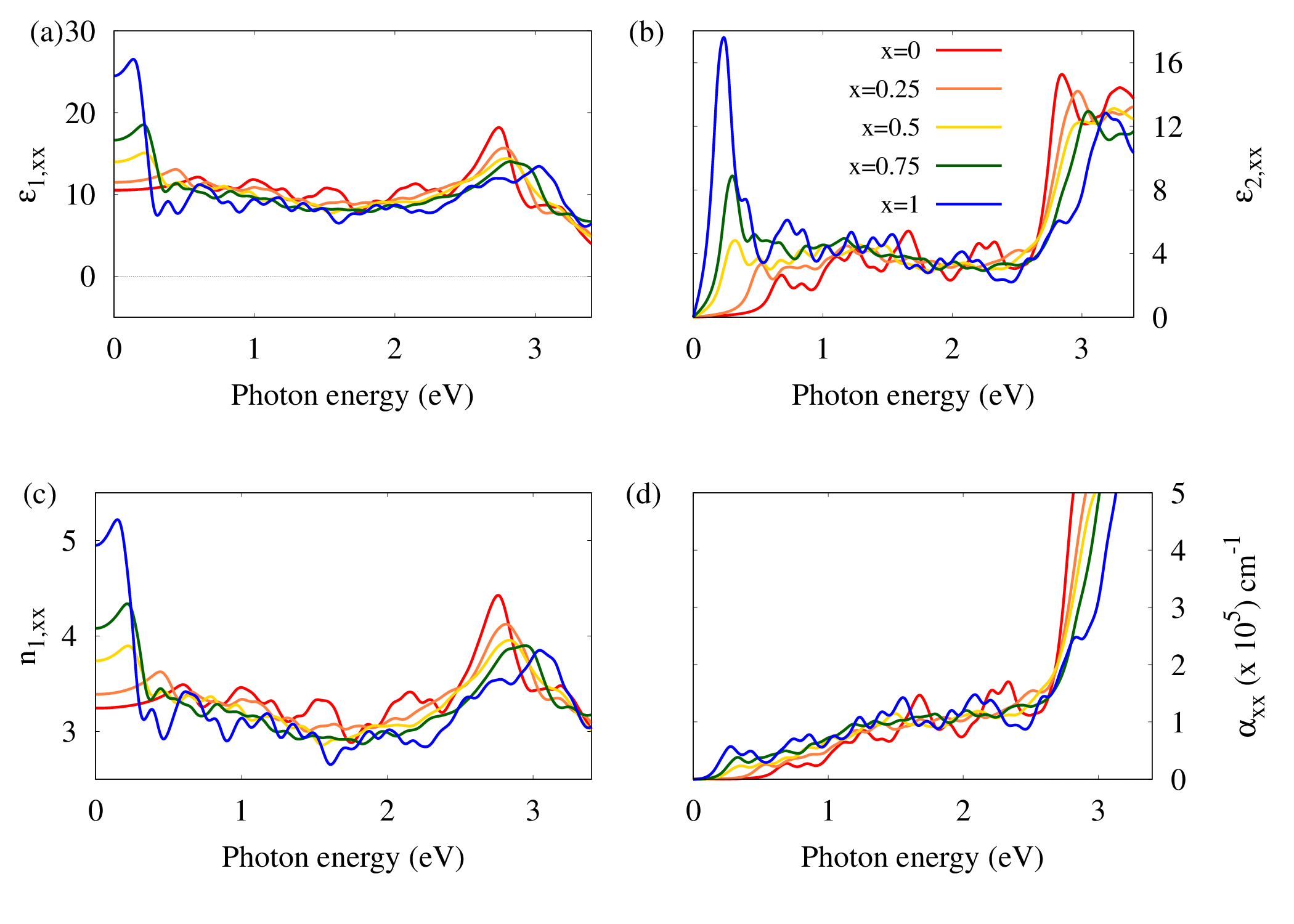}
\caption{Plots of a) Real part of dielectric constant b) Imaginary part of dielectric constant c) Real part of refractive index and d) Optical absorption coefficient of $\mathrm{KCuTe_{1-x}Se_{x}}$ at different concentrations in the xx-direction.}
\label{fig:kcutese_optical}
\end{figure*} 

The real and imaginary parts of the dielectric constant for x-, y- and z- polarized light are shown in Figures \ref{fig:kcutese_optical}, S3 and S4 (a) and (b). The dielectric constant of pure KCuTe and KCuSe in the hexagonal structure is isotropic for x- and y-polarized light but anisotropic for the z-polarized light, as expected. However, the asymmetry of the SQS at different alloy concentrations introduces an anisotropy in the dielectric constant for the x- and y-polarized light as well. This is evident from Tables \ref{table:static_epsilon_kcutese} and \ref{table:absorption_peaks}. The static dielectric constants, $\varepsilon_{1,xx}$ and $\varepsilon_{1,yy}$ are equal for KCuTe and KCuSe but are unequal for the SQS at different concentrations. The anistropy is also reflected in the differences in the positions of the first and second absorption peaks corresponding to the x- and y- polarized light of each SQS.

	\begin{table}[!hpbt]
	\caption{The static dielectric constant, $Re(\varepsilon)$ of $\mathrm{KCuTe_{1-x}Se_{x}}$ at different concentrations in the hexagonal structure. }
	\begin{tabular}{ cccc}
	\hline
	x & $\varepsilon_{1,xx}$(0) & $\varepsilon_{1,yy}$(0) & $\varepsilon_{1,zz}$(0) \\
	\hline
	0 & 10.53 & 10.53 & 5.42 \\
	0.25 & 11.49 & 11.35 & 5.46 \\
	0.5 & 13.99 & 14.46 & 5.49 \\
	0.75 & 16.65 & 16.68 & 5.55 \\
	1 & 24.50 & 24.50 & 5.61 \\
	\hline
	\end{tabular}
	\label{table:static_epsilon_kcutese}
	\end{table}

\begin{table}[!hpbt]
\caption{The major peaks in the imaginary dielectric constant, $\varepsilon_{2}$, in the visible energy spectrum. }
\begin{tabular}{ cccccc}
\hline
\multirow{2}{*}{x} & \multicolumn{3}{c}{$\mathrm{1^{st}}$ Absorption peak} & \multicolumn{2}{c}{$\mathrm{2^{nd}}$ Absorption peak} \\
 & xx & yy & zz & xx & yy \\
\hline
0 & 0.69 & 0.69 & 2.91 & 2.85 & 2.85 \\
0.25 & 0.53 & 0.53 & 2.93 & 2.97 & 2.97 \\
0.5 & 0.32 & 0.31 & 2.97 & 3.25 & 3.02 \\
0.75 & 0.30 & 0.29 & 3.00 & 3.05 & 3.06 \\
1 & 0.24 & 0.24 & 3.10 & 3.19 & 3.19 \\
\hline
\end{tabular}
\label{table:absorption_peaks}
\end{table}

In Figure \ref{fig:kcutese_optical}(b), the onset of light absorption occurs at a lower energy when the alloy concentration increases. This is directly related to the decrease in the bandgap of the alloy with increasing concentration. There is a corresponding shift in the position of the first absorption peak toward the lower energy with increasing alloy concentration (Table \ref{table:absorption_peaks}). The magnitude of $\varepsilon_{2,xx}$ at the first absorption peak increases significantly with increasing alloy concentration. The static dielectric constant, $\varepsilon_{1,xx}(0)$ and hence the refractive index, $n_{1,xx}(0)$ of the x-polarized light increases non-linearly with alloy concentration (Figures \ref{fig:kcutese_optical}(a), (c) and Table \ref{table:static_epsilon_kcutese}). Similar trends in the real and imaginary parts of the dielectric constant and refractive index are observed for the y-polarized light (Figure S3).

The imaginary part of the dielectric constant of the alloy for z-polarized light is shown in Figure S4 (b). The onset of absorption of z-polarized light by the alloy occurs at a very high energy compared to that of x- and y-polarised light. The magnitude of the first absorption peak for z-polarized light is significantly lower than that of the x- and y-polarized light. The real part of the dielectric constant and the refractive index are shown in Figures S4 (a) and (c). Unlike the x- and y-polarized light, the static dielectric constant and the refractive index for the z-polarized light does not vary with the change in the concentration of the alloy, indicating similar optical response of the alloy at all concentrations to z-polarized light.

The number of photogenerated electrons and holes in the semiconductor is essential for the performance of the PEC device and depends on the amount of incident light absorbed. When light of intensity $I_{0}$ is incident on a semiconductor, the intensity transmitted to depth $x$ is given by the Beer-Lambert relation. The absorption coefficient, $\alpha$, which is dependent on the wavelength of incident light, can be determined from estimates of the complex dielectric function $\varepsilon = \varepsilon_{1} + i\varepsilon_{2}$, according to: \(\alpha = \frac{\sqrt{2}\omega}{c}\sqrt{\sqrt{\varepsilon^{2}_{1}+\varepsilon^{2}_{2}}-\varepsilon_{1}}\), where, $\varepsilon_{1}$ is the real part and $\varepsilon_{2}$ is the imaginary part of the dielectric function.
The absorption coefficient of $\mathrm{KCuTe_{1-x}Se_{x}}$ at different concentrations is shown in Figure \ref{fig:kcutese_optical}. Because the bandgap of the alloy decreases with increase in the alloy concentration, the onset of absorption occurs at lower photon energy. The value of $\alpha$ within the visible spectrum is within 0.5 to 5$\times$$10^{5}$ $\mathrm{cm^{-1}}$, which makes it acceptable as a light absorber in PEC devices. 

\subsection{$\mathrm{KCuTe_{1-x}Se_{x}}$ as photocathode}

\begin{figure}[!htbp]
\centering
\includegraphics[width=0.45\textwidth]{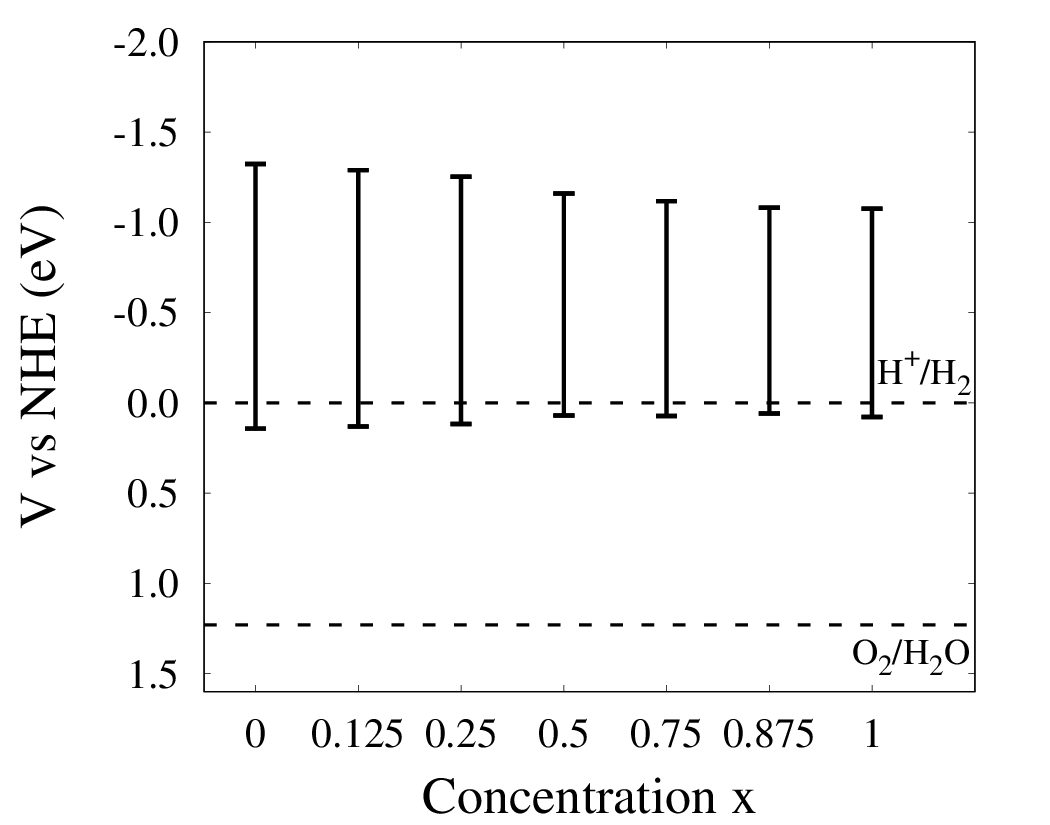}
\caption{Alignment of the CBE and VBE of $\mathrm{KCuTe_{1-x}Se_{x}}$ the water redox potentials with respect to NHE}
\label{fig:kcutese_band_alignment}
\end{figure}

Figure \ref{fig:kcutese_band_alignment} shows the position of the conduction band edge and the valence band edge with respect to the water oxidation and hydrogen reduction potentials obtained by the Butler-Ginley method. The valence band edge of the alloy at all concentrations is less positive than that of the water oxidation potential. As a result, the alloy cannot be used as a photoanode at any concentration. However, the conduction band edge of the alloy at all concentrations is more negative than the hydrogen reduction potential and can drive the proton reduction reaction at all concentrations. At the photocathode, electrons participate in the proton reduction reaction. The lower effective mass of electrons in the alloy implies a higher electron mobility. As a result, photogenerated electrons in the alloy are transferred to the electrode-electrolyte interface with fewer lost due to recombination compared to those of the pure compounds. This property further enhances its suitability as a photocathode.

\section{Conclusions}

From the theoretical investigation of $\mathrm{KCuTe_{1-x}Se_x}$ for photocathodes, we conclude that the volume deformation of KCuTe and KCuSe dominates the effects of charge exchange and structure relaxation and thus governs the energy of formation of $\mathrm{KCuTe_{1-x}Se_x}$ solid solutions. The temperature-composition phase diagram of the $\mathrm{KCuTe_{1-x}Se_{x}}$ alloy predicts a miscibility gap up to a critical peak temperature of 322 K. Beyond 322 K, the alloy forms a solid solution in the hexagonal structure between $0 \le x \le 1$. We then studied the electronic structure properties of $\mathrm{KCuTe_{1-x}Se_{x}}$ solid solutions using SQS in $\mathrm{P6_{3}/mmc}$. The electronic bandgap of the alloy varies linearly with concentration providing control over the photocathode bandgap to maximize the efficiency for a fixed photoanode. Despite satisfying the relative average orbital energy requirements of highly mismatched alloys, the relaxation of the structure competes with the deformation of the volume and the exchange of charges in $\mathrm{KCuTe_{1-x}Se_{x}}$, giving rise to a linear variation. Based on the direct nature of the bandgap and better tunability, lower electron effective mass, better light absorption in the visible spectrum, and the appropriate alignment of the conduction band with the proton reduction reaction, we propose that the $\mathrm{KCuTe_{1-x}Se_{x}}$ alloys are suitable materials for photocathodes.

\begin{acknowledgments}
We thankfully acknowledge Prof. Aftab Alam and Dr. Gurudayal Behera for valuable discussion. We acknowledge the Spacetime HPC facility at IIT Bombay. This work is supported by the IRCC with project code 07IR001 by IIT Bombay to DSM. 
\end{acknowledgments}

\bibliographystyle{unsrtnat} 
\bibliography{arinilit} 

\end{document}